# Spaser chains

E. S. Andrianov,[1,2] A. A. Pukhov,[1,2] A. V. Dorofeenko,[1,2] A. P. Vinogradov,[1,2] and A. A. Lisyansky[3]

[1]Moscow Institute of Physics and Technology, 9 Institutskiy per., Dolgoprudniy 141700, Moscow Reg., Russia

[2]Institute for Theoretical and Applied Electromagnetics, 13 Izhorskaya, Moscow 125412, Russia

[3]Department of Physics, Queens College of the City University of New York, Flushing, NY 11367, USA

**Abstract**

We show that depending on the values of the coupling constants, two different scenarios for the stationary behavior of a chain of interacting spasers may be realized: (1) all the spasers are synchronized and oscillate with a unique phase and (2) a nonlinear autowave travels along the chain. In the latter scenario, the traveling wave is *harmonic* unlike excitations in other known nonlinear systems. The amplitude of this wave is determined by pumping and the wavenumber is determined by the coupling constants. Due to the nonlinear nature of the system, any initial distribution of spasers' states evolves into one of these steady states.

## I. INTRODUCTION

In the last decade, quantum nanoplasmonics has experienced explosive growth due to numerous revolutionary applications in optics and optoelectronics (see Refs. 1-4 and references therein). The rapid emergence of this field reflects the increasing demands of nanotechnology. Nanoplasmonics utilizes outstanding optical properties of metal nanoparticles (NPs) which allow the electromagnetic field to be concentrated on a subwavelength scale. A combination of a nanoscale active medium with the population inversion, e.g. a quantum dot (QD), with a metal NP results in the emergence of a nanoplasmonic counterpart of the laser – surface plasmon amplification by stimulated emission of radiation (spaser) first proposed by Bergman and Stockman[5] and realized experimentally by Noginov at al.[6] Schematically, the spaser is a system of inversely excited two-level QDs surrounding metal NPs.[5, 7] The principles of operation of the spaser are analogous to those for the laser with the role of photons played by surface plasmons (SPs) localized at a NP that serves as the resonator.[5, 7] The role of photons is played by SPs, which are localized at the NP.[5, 7-9] Confining SPs to the NP resembles a resonator. The spaser generates and amplifies the near field of the NP. SP amplification occurs due to nonradiative energy transfer from the QD to the NP. This process originates from the dipole-dipole (or any other near field[10]) interaction between the QD and the plasmonic NP. This physical mechanism has high efficiency because the probability of exciting the SP excitation is $\sim (kr)^{-3}$ larger than the probability of radiative emission,[11] where $r$ is the distance between the centers of the NP and the QD and $k$ is an optical wavenumber in vacuum. The generation of a large number of SPs leads to the induced emission of the QD into the plasmonic mode and to the development of generation of plasmons. Thus, the excitation of the plasmonic mode is provided by pumping through the excited QD. This process is inhibited by losses in the NP, which together with pumping results in undamped stationary oscillations of the spaser dipole moment.

A promising application of spasers is their use as inclusions in plasmonic metamaterials – composites with negative dielectric permittivity and magnetic permeability (for review see Refs. 12 and 13). The inhibiting factor for applications of metamaterials is a substantial Joule loss due to the interacting of light with metal inclusions that fill metamaterials. This loss can be compensated by using metal inclusions for building spasers inside metamaterials.[14, 15] Using spasers for the loss compensation is perspective but not straightforward. The problem is that below the generation threshold, the spaser is an autonomous (self-oscillating) system. Its dipole moment is excited not only by the external field but also by the radiation produced by the QD. Moreover, the dipole moment of the spaser's NP oscillates even in the absence of the external field. The frequency of this self-oscillation is determined by the plasmon frequency, the transition frequency of the QD and characteristic times of relaxation of excitations in NP and QD.[9] Thus, a spaser is not necessarily synchronized with the external field. Nevertheless, as has been shown in Ref. 16, there is a domain (the Arnold tongue) of values of the external wave amplitude and the frequency detuning at which the spaser becomes synchronous with the driving wave and even exact compensation of losses can be achieved.[17] This may lead to the creation of lossless metamaterials.

If spasers are used as the gain medium for loss compensation, one must understand the functioning of a system of interacting spasers distributed regularly or randomly in a dielectric matrix. In the present paper, we study a regular linear chain of interacting spasers. We demonstrate that, depending on the parameters of the system, two different scenarios may be realized in such a system.

There are several papers devoted to the numerical simulation of electromagnetic wave traveling through a periodic system of spasers.[18-21] Unlike our study, in which a nonlinear regime of the SP generation in spasers (spasing) is considered, in Refs. 18-21 a linear regime of interaction among spasers and between spasers and external optical wave has been studied. This linear regime takes place during short probe pulses when the population inversion is not depleted markedly, or when the spasers operate below the spasing threshold. This restriction allows one to imply that the system can be described by the effective permittivity. Our study deals with the nonlinear regime when the population inversion is above the spasing threshold.

The rest of the paper is organized as follows. In Section II, we briefly discuss main equations that describe operations of spasers. In Section III, we consider oscillations of two coupled spasers. In Sections IV and V, we find solutions and analyze their stability for a model of a chain of spasers in which only neighboring NPs are coupled. Section VI, follows with a more general model in which the interaction of NPs with nearest QDs is included. This model is generalized further in Section VII, to take into account the coupling of QDs that belong to neighboring spasers. Finally, Section VIII summarizes the paper.

## II. OPERATION OF A SPASER. MAIN EQUATIONS

The simplest spaser consists of a two-level QD of radius $r_{TLS}$, which is positioned at a distance $r$ from a metallic NP of radius $r_{NP}$. The whole system is immersed into a solid dielectric or semiconductor matrix with dielectric permittivity $\varepsilon_M$.[5] Below, we discuss the excitation of the main, dipole, mode of the SP at a frequency $\omega_{SP}$.

Following the theory of a usual one-atom single-mode laser,[22] the dynamics of the self-oscillating spaser can be described by the following Hamiltonian:

$$\hat{H} = \hat{H}_{SP} + \hat{H}_{TLS} + \hat{V} + \hat{\Gamma}, \qquad (1)$$

where

$$\hat{H}_{SP} = \hbar\omega_{SP}\hat{\tilde{a}}^\dagger\hat{\tilde{a}}, \qquad (2)$$

$$\hat{H}_{TLS} = \hbar\omega_{TLS}\hat{\tilde{\sigma}}^\dagger\hat{\tilde{\sigma}} \qquad (3)$$

describe the energy of the electromagnetic field of the SP and of the QD, respectively,[5, 12, 23] the operator $\hat{V} = -\hat{\boldsymbol{\mu}}_{TLS} \cdot \hat{\mathbf{E}}_{NP}$ determines the coupling between the QD and NP, the operator $\hat{\Gamma}$ is responsible for relaxation and pumping,[23] $\hat{a}(t)$ is the Bose operator of annihilation of the dipole SP, $\hat{\sigma} = |g\rangle\langle e|$ is the operator of the transition between the excited $|e\rangle$ and ground $|g\rangle$ states of the QD, $\hat{\boldsymbol{\mu}}_{TLS} = \boldsymbol{\mu}_{TLS}\left(\hat{\sigma}(t) + \hat{\sigma}^\dagger(t)\right)$ is the dipole moment of the QD, and $\boldsymbol{\mu}_{TLS} = \langle e|e\mathbf{r}|g\rangle$ is its off-diagonal matrix element.. The operator of the electric field has the form

$$\hat{\mathbf{E}}_{NP}(\mathbf{r},t) = -A\nabla\varphi(\mathbf{r})\left(\hat{a}^\dagger + \hat{a}\right), \qquad (4)$$

where $A = (4\pi\hbar s / \varepsilon_d s')^{1/2}$, $s' = d(\mathrm{Re}[s(\omega)])/d\omega\big|_{\omega=\omega_n}$, $s(\omega) = [1 - \varepsilon_{NP}(\omega)/\varepsilon_M(\omega)]^{-1}$, and $\varphi$ is the potential of the SP field. Since the SP wavelength, $\lambda_{SP}$, is much smaller than the optical wavelength in vacuum, $\lambda$, we can use the quasistatic approximation, in which the far fields are neglected. Thus, one can find the frequency of the plasmonic resonance from the condition of existence of the nontrivial solution of the Laplace equation

$$\nabla\varepsilon(\mathbf{r})\nabla\varphi(\mathbf{r}) = 0,$$

in which $\varepsilon(\mathbf{r})$ equals $\varepsilon_{NP}$ inside the NP and $\varepsilon_M$ outside the NP in the dielectric matrix. To separate the electromagnetic properties and the geometrical factor of the NP, it is convenient to represent the permittivity in the form[5] $\varepsilon(\mathbf{r}) = (\varepsilon_{NP} - \varepsilon_M)(\Theta(\mathbf{r}) - s)$, where the step function $\Theta(\mathbf{r})$ defines the geometry of the problem: it is equal to zero in the dielectric matrix and equal to unity inside the NP, whereas $s = (1 - \varepsilon_{NP}/\varepsilon_M)^{-1}$ only depends on constitutive properties of the system. By using such notation the Laplace equation may be written in the form[8,9]

$$\nabla\Theta(\mathbf{r})\nabla\varphi(\mathbf{r}) = s\nabla^2\varphi(\mathbf{r}), \qquad (5)$$

which allows one to find the eigenvalues $s_n$ and eigenfunctions $\varphi_n(\mathbf{r})$, $n = 1, 2, ...$ In the particular case of a spherical NP, the $n$-th plasmonic multipole has the resonance at $\varepsilon_{NP} = -\varepsilon_M(n+1)/n$ and $s_n = n/(2n+1)$.[11,12,24]

By taking into account the permittivity dispersion, $\varepsilon_{NP}(\omega)$, the eigenfrequency $\Omega_n$ is determined from the condition $s(\Omega_n) = [1 - \varepsilon_{NP}(\Omega_n)/\varepsilon_M]^{-1} = s_n$. In the general case, due to the Joule losses in the metallic NP, the eigenfrequency has an imaginary part: $\Omega_n = \omega_n - i\gamma_n$. In addition, there are radiation losses of the NP due to the oscillation of a SP multipole. However, for small NPs (~30 nm). Joule losses far exceed the loss due to radiation.[12, 25] This means that the spaser mainly generates the near field. For large NPs (~100 nm) the loss due to radiation dominates. This is the regime of the so-called dipole laser.[26] In this paper, we assume that the size of the NP is small and disregard the radiative losses.

Below, for the quantization of the electromagnetic field we take into account the dispersion but neglect dissipation[27]. The losses and pumping are introduced into the final equations phenomenologically (see Ref. 26 for details).

Assuming that the frequency of the QD transition is close to the frequency of the plasmon resonance, $\omega_{SP} \approx \omega_{TLS}$, we can represent time dependencies of $\hat{\tilde{a}}(t)$ and $\hat{\tilde{\sigma}}(t)$ in the the form $\hat{\tilde{a}}(t) \equiv \hat{a}(t)e^{-i\omega t}$ and $\hat{\tilde{\sigma}}(t) \equiv \hat{\sigma}(t)e^{-i\omega t}$, where $\hat{a}(t)$ and $\hat{\sigma}(t)$ are slowly varying operators and $\omega$ is the spaser's self-oscillation frequency which we seek. In the approximation of the rotating wave[22] we can neglect fast oscillating terms, which are proportional to $\sim e^{\pm 2i\omega t}$. Then the coupling operator can be written in the Jaynes-Cummings form[22]

$$\hat{V} = \hbar\Omega_R(\hat{a}^\dagger\hat{\sigma} + \hat{\sigma}^\dagger\hat{a}). \quad (6)$$

where $\Omega_R$ is the Rabi frequency.

The commutation relations for operators $\hat{a}(t)$ and $\hat{\sigma}(t)$ are standard: $\left[\hat{a},\hat{a}^\dagger\right] = \hat{1}$ and $\left[\hat{\sigma}^\dagger,\hat{\sigma}\right] = \hat{D}$, where the operator $\hat{D} = \left[\hat{\sigma}^\dagger,\hat{\sigma}\right] = \hat{n}_e - \hat{n}_g$ describes the population inversion of the ground $\hat{n}_g = |g\rangle\langle g|$ and excited states $\hat{n}_e = |e\rangle\langle e|$, $\hat{n}_g + \hat{n}_e = \hat{1}$, of the QD. Using the Hamiltonian, Eqs. (1)-(6), we obtain Heisenberg equations of motion for operators $\hat{a}(t)$, $\hat{\sigma}(t)$, and $\hat{D}(t)$:[8, 16, 26]

$$\dot{\hat{D}} = 2i\Omega_R(\hat{a}^\dagger\hat{\sigma} - \hat{\sigma}^\dagger\hat{a}) - \tau_D^{-1}\left(\hat{D} - \hat{D}_0\right), \quad (7)$$

$$\dot{\hat{\sigma}} = \left(i\delta - \tau_\sigma^{-1}\right)\hat{\sigma} + i\Omega_R\hat{a}\hat{D}, \quad (8)$$

$$\dot{\hat{a}} = \left(i\Delta - \tau_a^{-1}\right)\hat{a} - i\Omega_R\hat{\sigma}, \quad (9)$$

where $\delta = \omega - \omega_{TLS}$ and $\Delta = \omega - \omega_{SP}$ are frequency detunings. The terms with relaxation times $\propto \tau_D^{-1}$, $\tau_\sigma^{-1}$, and $\tau_a^{-1}$ are introduced to account for the relaxation processes of the respective quantities. These relaxation times can be evaluated in terms of macroscopic properties of QD and NP. For typical QD and silver NP these values are:[28-32] $\tau_a^{-1} \sim 10^{14}\,s^{-1}$, $\tau_\sigma^{-1} \sim 10^{11}\,s^{-1}$, $\tau_D^{-1} \sim 10^{13}\,s^{-1}$. The operator $\hat{D}_0$ describes pumping. The population inversion operator $\hat{D}_0$ plays the role of the operator $\hat{D}$ in the regime of absence of generation. In other words, without generation $\{\hat{D} = \hat{D}_0, \hat{a} = \hat{\sigma} = \hat{0}\}$ is a stable fixed point of Eqs. (7)-(9). In the case when generation exists, this fix point becomes unstable.[22, 23]

Below we neglect quantum fluctuations and correlations and consider $\hat{a}(t)$, $\hat{\sigma}(t)$, and $\hat{D}(t)$ as complex valued quantities (*c*-numbers), so that we can use complex conjugation instead of the Hermitian conjugation.[5, 26, 33, 34] The population inversion $D(t)$ must be a real valued quantity because the respective operator is Hermitian. The quantities $\sigma(t)$ and $a(t)$ are the complex amplitudes of the dipole oscillations of the QD and SP, respectively.

As has been shown in Refs. 5 and 26, above the threshold pump level,

$$D_{th} = \frac{1+\Delta^2\tau_a^2}{\Omega_R^2\tau_a\tau_\sigma}, \tag{10}$$

the spaser oscillates with the frequency

$$\omega_a = \frac{\omega_{SP}\tau_a + \omega_{TLS}\tau_\sigma}{\tau_a + \tau_\sigma}, \tag{11}$$

and Eqs. (7)-(9) have stationary solutions

$$D = D_{th}, \tag{12}$$

$$\sigma = \frac{a}{\Omega_R}\left(\Delta + i\tau_a^{-1}\right), \tag{13}$$

$$a = \frac{e^{i\psi}}{2}\left(\frac{(D_0 - D_{th})\tau_a}{\tau_D}\right)^{\frac{1}{2}}, \tag{14}$$

where $\psi$ is an undetermined phase shift, which value is immaterial. The frequency of the spaser's self-oscillations is always between the frequencies of the plasmon resonance and the QD transition. This is similar to the frequency pulling, which is the well-known phenomenon in lasers. The square-root dependence of the amplitudes of dipole moments of metal NPs and QDs on pumping corresponds to the standard Hopf bifurcation at the creation of the limit cycle. It is important to note that in our approximation, the arising cycle has a strictly circular shape, so that even though spaser's self-oscillations are nonlinear, they are harmonic. As we show below, the nonlinear autowave of the spasing propagation is also harmonic.

### III. SYNCHRONIZATION OF TWO SPASERS

The spaser, as a self-oscillating system, exhibits properties common to such systems. One of them has been known as far back as to Huygens is synchronization.[35] The synchronization may occur either under an action of the external force or due to the interaction of two self-oscillating systems. In both cases, one observes phase locking when either the system driven by an external harmonic field starts oscillating with the frequency of this field[16] or when two coupled self-oscillating systems start oscillating in phase with the same frequency. The case of the spaser synchronization by the external wave was considered in Ref. 16. In this section, we discuss synchronization of coupled spasers.

To describe behavior of two coupled spasers we should generalized the system of equations (7)-(9) by introducing coupling terms:

$$\dot{\hat{D}}_1 = 2i\Omega_{R1}(\hat{a}_1^\dagger \hat{\sigma}_1 - \hat{\sigma}_1^\dagger \hat{a}_1) + 2i\Omega_{NP-TLS}(\hat{a}_2^\dagger \hat{\sigma}_1 - \hat{\sigma}_1^\dagger \hat{a}_2) + 2i\Omega_{TLS-TLS}(\hat{\sigma}_2^\dagger \hat{\sigma}_1 - \hat{\sigma}_1^\dagger \hat{\sigma}_2) - \tau_{D1}^{-1}(\hat{D}_1 - \hat{D}_{01}), \quad (15)$$

$$\dot{\hat{D}}_2 = 2i\Omega_{R2}(\hat{a}_2^\dagger \hat{\sigma}_2 - \hat{\sigma}_2^\dagger \hat{a}_2) + 2i\Omega_{NP-TLS}(\hat{a}_1^\dagger \hat{\sigma}_2 - \hat{\sigma}_2^\dagger \hat{a}_1) + 2i\Omega_{TLS-TLS}(\hat{\sigma}_2^\dagger \hat{\sigma}_1 - \hat{\sigma}_1^\dagger \hat{\sigma}_2) - \tau_{D2}^{-1}(\hat{D}_2 - \hat{D}_{02}), \quad (16)$$

$$\dot{\hat{\sigma}}_1 = (i\delta - \tau_\sigma^{-1})\hat{\sigma}_1 + i\Omega_{R1}\hat{a}_1\hat{D}_1 + i\Omega_{NP-TLS}\hat{a}_2\hat{D}_1 + i\Omega_{TLS-TLS}\hat{\sigma}_2\hat{D}_1, \quad (17)$$

$$\dot{\hat{\sigma}}_2 = (i\delta - \tau_\sigma^{-1})\hat{\sigma}_2 + i\Omega_{R2}\hat{a}_2\hat{D}_2 + i\Omega_{NP-TLS}\hat{a}_1\hat{D}_2 + i\Omega_{TLS-TLS}\hat{\sigma}_1\hat{D}_2,, \quad (18)$$

$$\dot{\hat{a}}_1 = (i\Delta - \tau_a^{-1})\hat{a}_1 - i\Omega_{R1}\hat{\sigma}_1 - i\Omega_{NP-NP}\hat{a}_2 - i\Omega_{NP-TLS}\hat{\sigma}_2, \quad (19)$$

$$\dot{\hat{a}}_2 = (i\Delta - \tau_a^{-1})\hat{a}_2 - i\Omega_{R2}\hat{\sigma}_2 - i\Omega_{NP-NP}\hat{a}_1 - i\Omega_{NP-TLS}\hat{\sigma}_1. \quad (20)$$

Indexes 1 and 2 label the quantities corresponding to the first and to second spasers, respectively,

$$\Omega_{NP-NP} = \frac{\boldsymbol{\mu}_{NP1}\cdot\boldsymbol{\mu}_{NP2} - 3(\boldsymbol{\mu}_{NP1}\cdot\mathbf{e}_{NP-NP})(\boldsymbol{\mu}_{NP2}\cdot\mathbf{e}_{NP-NP})}{\hbar r_{NP-NP}^3}$$

is a coupling constant between neighboring NP,

$$\Omega_{NP-TLS} = \frac{\boldsymbol{\mu}_{NP1} \cdot \boldsymbol{\mu}_{TLS2} - 3(\boldsymbol{\mu}_{NP1} \cdot \mathbf{e}_{NP-TLS})(\boldsymbol{\mu}_{TLS2} \cdot \mathbf{e}_{NP-TLS})}{\hbar r_{NP-TLS}^3}$$

$$= \frac{\boldsymbol{\mu}_{NP2} \cdot \boldsymbol{\mu}_{TLS1} - 3(\boldsymbol{\mu}_{NP2} \cdot \mathbf{e}_{NP-TLS})(\boldsymbol{\mu}_{TLS1} \cdot \mathbf{e}_{NP-TLS})}{\hbar r_{NP-TLS}^3}$$

is a coupling constant between neighboring QD and NP,

$$\Omega_{TLS-TLS} = \frac{\boldsymbol{\mu}_{TLS1} \cdot \boldsymbol{\mu}_{TLS2} - 3(\boldsymbol{\mu}_{TLS1} \cdot \mathbf{e}_{TLS-TLS})(\boldsymbol{\mu}_{TLS2} \cdot \mathbf{e}_{TLS-TLS})}{\hbar r_{TLS-TLS}^3}$$

is a coupling constant between neighboring QDs, $\mathbf{e}_{NP-NP}$, $\mathbf{e}_{NP-TLS}$, and $\mathbf{e}_{TLS-TLS}$ are unit vectors pointing from the NP to the neighboring NP, from the NP to the QD, and from the QD to the neighboring QD, respectively, $r_{NP-NP}$, $r_{NP-TLS}$, and $r_{TLS-TLS}$ are the corresponding distances. These coupling constants have the dimension of frequency. For typical QD and NP at optical frequencies, $\sim 5 \cdot 10^{15} s^{-1}$, supposing that $r_{NP-NP}, r_{NP-TLS}, r_{TLS-TLS} \sim 50 nm$, we obtain $\Omega_{NP-NP} \sim 10^{13} s^{-1}$, $\Omega_{NP-TLS} \sim 10^{12} s^{-1}$, $\Omega_{TLS-TLS} \sim 10^{11} s^{-1}$.

The values of the coupling constants depend on the orientation of corresponding dipoles. In particular, if the dipoles are parallel, then $\Omega_{1-2} = \gamma_i (\boldsymbol{\mu}_1 \cdot \boldsymbol{\mu}_2)/(\hbar r_{12}^3)$ with $\gamma_1 = -2$ for the longitudinal and $\gamma_{2,3} = 1$ for transverse modes. Below we imply that the $\gamma$ factor is included in the definition of the coupling constants.

One can show that for $D_0 < D_{th}$, where

$$D_{th} = \frac{1 + (\Omega_{NP-NP}^{eff} \tau_\sigma)^2}{(\Omega_R + 2\Omega_{NP-TLS})^2 \tau_a \tau_\sigma}, \tag{21}$$

only the trivial solution, $a_{1,2} = 0$, $\sigma_{1,2} = 0$ and $D_{1,2} = D_{01,2}$, satisfies the system of equations (15)-(20). In Eq. (21) we introduce the notation $\Omega_{NP-NP}^{eff} = 2\Omega_{NP-NP} \tau_a / (\tau_a + \tau_\sigma)$. For $D_0 > D_{th}$, by the direct substitution one can verify that if $\Omega_{NP-TLS} > 0$, then

$$\hat{a}_1 = \hat{a}_2, \quad \hat{\sigma}_1 = \hat{\sigma}_2, \quad \hat{D}_1 = \hat{D}_2, \tag{22a}$$

is the stationary solution to Eqs. (15)-(20). For $\Omega_{NP-TLS} < 0$ the stationary solution is

$$\hat{a}_1 = -\hat{a}_2, \quad \hat{\sigma}_1 = -\hat{\sigma}_2, \quad \hat{D}_1 = \hat{D}_2. \tag{22b}$$

Thus, when the pumping exceeds the threshold value (21), a couple of identical spasers behaves like a single spaser for which the spasing arise with the frequency

$$\omega = \Omega_{NP-NP}^{eff} + \frac{\omega_{SP}\tau_a + \omega_{TLS}\tau_\sigma}{\tau_a + \tau_\sigma}. \quad (23)$$

In this case, the phase locking occurs, so that phase difference $\Delta\varphi$ between the oscillations is equal to zero and $\pi$, if $\Omega_{NP-TLS} > 0$ and $\Omega_{NP-TLS} < 0$, respectively. This result is confirmed by computer simulation In time domain that shows that the system arrives at the stationary state given by Eqs. (21)-(23) independent of initial conditions.

## IV. MODEL HAMILTONIAN OF A CHAIN OF SPASERS

In this section we discuss a chain of spasers schematically shown in Fig. 1. As mentioned above (see also Refs. 5 and 26), the main contribution to the interaction between spaser's members, NPs and QDs, is due to near fields. At the first step, considering a linear chain of spasers we take into account the dipole-dipole interaction, $\Omega_{NP-NP}$, of NPs of the nearest-neighbor spasers disregarding far-field interaction as well as the influence of neighboring NP on QD, which is considered in Sec. VI.

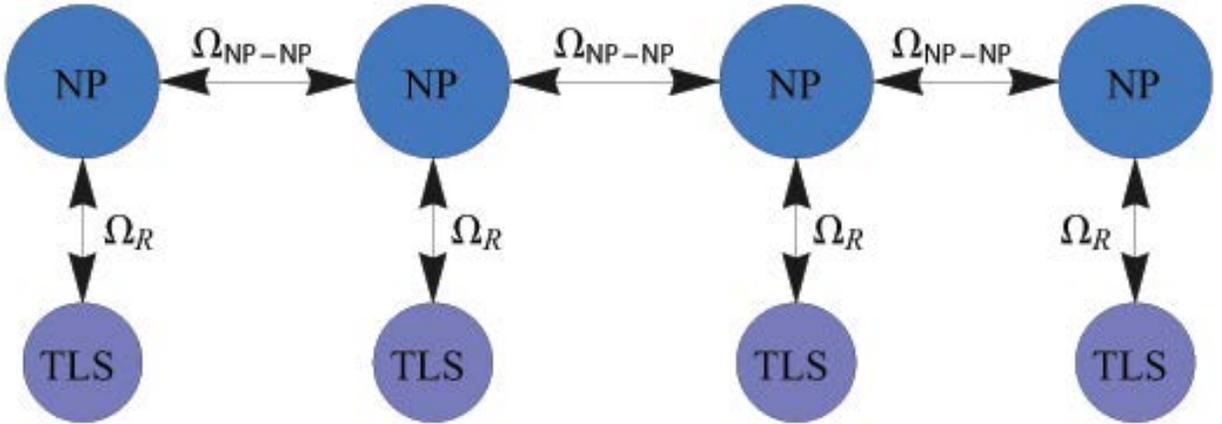

Fig. 1. (Color online) Schematic of the chain of spasers, in which NPs of neighboring interact with each other.

In the framework of the slowly varying envelope approximation, $\hat{a}_n(t)\exp(-i\omega t)$, $\hat{\sigma}_n(t)\exp(-i\omega t)$, the Hamiltonian of the chain of spasers with interacting NPs can be written as

$$\hat{H} = \sum_n \left( \hbar\omega_{SP}\hat{a}_n^+\hat{a}_n + \hbar\omega_{TLS}\hat{\sigma}_n^+\hat{\sigma}_n + \hbar\Omega_R\left(\hat{a}_n^+\hat{\sigma}_n + \hat{a}_n\hat{\sigma}_n^+\right) \right.$$
$$\left. + \frac{1}{2}\hbar\Omega_{NP-NP}\left(\hat{a}_n^+\hat{a}_{n-1} + \hat{a}_n\hat{a}_{n-1}^+ + \hat{a}_n^+\hat{a}_{n+1} + \hat{a}_n\hat{a}_{n+1}^+\right) \right), \quad (24)$$

where the summation is performed over the pasers. The Heisenberg equations for the *n*-th spaser can be written as

$$\dot{\hat{D}}_n = 2i\Omega_R(\hat{a}_n^+\hat{\sigma}_n - \hat{\sigma}_n^+\hat{a}_n) - \tau_D^{-1}\left(\hat{D}_n - \hat{D}_{0n}\right), \quad (25)$$

$$\dot{\hat{\sigma}}_n = \left(i\delta - \tau_\sigma^{-1}\right)\hat{\sigma}_n + i\Omega_R\hat{a}_n\hat{D}_n, \quad (26)$$

$$\dot{\hat{a}}_n = \left(i\Delta - \tau_a^{-1}\right)\hat{a}_n - i\Omega_R\hat{\sigma}_n - i\Omega_{NP-NP}(\hat{a}_{n-1} + \hat{a}_{n+1}). \quad (27)$$

In Eq. (27) the last term is responsible for the spaser coupling.

Next, we replace operators by *c*-numbers and look for stationary solutions by considering equations:

$$2i\Omega_R(a_n^*\sigma_n - \sigma_n^*a_n) - \tau_D^{-1}\left(D_n - D_{0n}\right) = 0, \quad (28)$$

$$\left(i\delta - \tau_\sigma^{-1}\right)\sigma_n + i\Omega_R a_n D_n = 0, \quad (29)$$

$$\left(i\Delta - \tau_a^{-1}\right)a_n - i\Omega_R\sigma_n - i\Omega_{NP-NP}(a_{n-1} + a_{n+1}) = 0. \quad (30)$$

We assume that the solution of Eqs. (28)-(30) is a plane travelling wave with a slowly varying amplitude: $a_n = a_{n,k}\exp(ikx)$ and $\sigma_n = \sigma_{n,k}\exp(ikx)$. Since *x* takes discrete values, $x = nb$, where *b* is the distance between nearest-neighbor spasers, we obtain

$$a_{n+1,k} = a_{n,k}\exp(ikb), \quad a_{n-1} = a_n\exp(-ikb). \quad (31)$$

Substituting these expressions into system (28)-(30) we obtain the solution which is identical to the solution to the system of equations (7)-(9) for a single spaser (see for details Refs. 16 and 26) with the replacement

$$\Delta \rightarrow \Delta(k) - 2\Omega_{NP-NP}\cos kb. \quad (32)$$

Note, that in this case, $\Delta(k) = \omega_k - \omega_a$. The dispersion equation $\tau_a\Delta = -\tau_\sigma\delta$ of a solitary spaser,[16, 26] which is resulted from Eqs. (7)-(9), now takes the form

$$\tau_a\left(\Delta(k) - 2\Omega_{NP-NP}\cos kb\right) = -\tau_\sigma\delta. \quad (33)$$

The solution to Eq. (33) gives us the dispersion equation

$$\omega_k = \omega_a + \Omega^{eff}_{NP-NP} \cos kb, \qquad (34)$$

where the generation frequency of a single spaser, $\omega_a$, is given by Eq. (11),[26] and $\omega_0$ is the frequency of SPs.

Thus, even though the chain of spasers is a non-linear system, *harmonic* waves can travel in it. The dispersion equation for these waves is similar to the dispersion equation for the wave of dipole moments traveling along the linear system of lossless chain of spherical particles in the absence of QDs.[11, 36, 37] Indeed, in the latter case, the dispersion equation for one longitudinal and two transverse modes has the form

$$\omega(k) = \omega_{SP} + \left(\gamma_i \omega_1^2 / \omega_{SP}\right) \cos kb. \qquad (35)$$

As above, $\gamma_1 = -2$ for the longitudinal and $\gamma_{2,3} = 1$ for transverse modes. In Ref. 2 it has been shown that $\omega_1^2 = r_{NP}^3 \omega_{pl}^2 / 3\hbar b^3$, where $\omega_{pl}$ is the plasmon frequency in the Drude type of dispersion $\varepsilon_{NP} = 1 - \omega_{pl}^2 / \omega^2$. Taking into account that $|\mu_{NP}| = \sqrt{3\hbar r_{NP}^3 / (\partial \operatorname{Re} \varepsilon_{NP} / \partial \omega)}$ and the assumption[2] that the surface plasmon resonance occurs at $\omega_{SP} = \omega_{pl} / \sqrt{3}$, the factor in Eq. (34) can be recast as

$$\Omega^{eff}_{NP-NP} = 2\Omega_{NP-NP} \frac{\tau_a}{\tau_a + \tau_\sigma} = \frac{2\tau_a}{\tau_a + \tau_\sigma} \gamma_i \frac{\mu_{NP}^2}{\hbar b^3} = \frac{\tau_a}{\tau_a + \tau_\sigma} \gamma_i \frac{\omega_1^2}{\omega_{SP}},$$

which up to the factor $\tau_a / (\tau_a + \tau_\sigma)$ coincides with the factor in Eq. (35). In the case of absence of the QD, we can put $\tau_\sigma = 0$ and arrive at exact coincidence.

Thus, we obtain that the dispersion equations for the harmonic waves traveling along a chain of plasmonic NPs and a chain of nonlinear spasers are nearly the same. Setting $\tau_\sigma$ equal to zero, makes $\omega_a$ equal to $\omega_{SP}$ and $\Omega^{eff}_{NP-NP}$ in Eq. (34) equal to the factor $\omega_1^2 / \omega_{SP}$ in Eq. (35). However, there is a principal difference between the waves in these systems. In the linear system, the wave amplitude can take an arbitrary value, while in the nonlinear case the amplitude is determined by the pumping:

$$a_{n,k} = \frac{1}{2} \exp(i\varphi) \sqrt{\left[D_0 - \left(1 + \left(\Omega^{eff}_{NP-NP} \tau_\sigma\right)^2 \cos^2 kb\right) \Big/ \Omega_R^2 \tau_a \tau_\sigma\right] \frac{\tau_a}{\tau_\sigma}}. \qquad (36)$$

In the linear case, we can have a superposition of waves: two or more waves can travel simultaneously. Their amplitudes and wavenumbers are determined by initial conditions. That is not the case for the nonlinear system. In the nonlinear system regardless of the initial conditions, the threshold value $D_{th}$ of the inversion of population, for which the wave can arise, depends on the wavenumber $k$:

$$D_{th} = \frac{1+(\Delta - 2\Omega_{NP-NP}\cos kb)^2 \tau_a^2}{\Omega_R^2 \tau_a \tau_\sigma} = \frac{1+\left(\Omega_{NP-NP}^{eff}\tau_\sigma\right)^2 \cos^2 kb}{\Omega_R^2 \tau_a \tau_\sigma}. \tag{37}$$

In terms of the frequency this relation can be written as

$$D_{th} = \frac{1+(\omega_a - \omega_k)^2 \tau_a \tau_\sigma}{\Omega_R^2 \tau_a \tau_\sigma}. \tag{38}$$

The value of $D_{th}$ has minimum at $\omega_k = \omega_a$, the corresponding wavenumber is equal to $k = \pi/2b$. The wave with this frequency and wavenumber is the first to be generated. We can assume that only this wave survives and the other waves transfer their energy to this one. To confirm the assumption let us consider how a small disturbance affects a nonlinear harmonic wave.

## V. ANALYSIS OF STABILITY

The solution to Eqs. (28)-(30) has the form

$$D_{st}(k) = D_{th}(k) = \frac{1+\left(\Omega_{NP-NP}^{eff}\tau_\sigma\right)^2 \cos^2 kb}{\Omega_R^2 \tau_a \tau_\sigma}, \tag{39}$$

$$a_{st}(k) = \frac{e^{i\psi}}{2}\sqrt{\left(D_0 - D_{th}(k)\right)\frac{\tau_a}{\tau_D}}, \tag{40}$$

$$\sigma_{st} = a_{st}\left(\Delta - 2\Omega_{NP-NP}\cos kb + i/\tau_a\right)/\Omega_R = a_{st}\left(i - \Omega_{NP-NP}^{eff}\tau_\sigma \cos kb\right)/\tau_a \Omega_R, \tag{41}$$

with $k(\omega)$ determined by Eq. (35).

To verify the stability of this solution we add to it a disturbance of the form $\exp(\Lambda_\chi t + i\chi x)$. Here $x = bn$, $n$ is a spaser's number, $\chi$ is the wavenumber of the disturbance, $\Lambda_\chi$ is the increment of growth of this disturbance. We consider the initial autowave as stable if for any $\chi$ we obtain $\mathrm{Re}\,\Lambda_\chi < 0$. To proceed it is convenient to rewrite system of equations (25)-(27) in a vector form

$$\dot{\mathbf{u}} = \mathbf{f}(\mathbf{u}), \tag{42}$$

where

$$\mathbf{u} = \begin{pmatrix} A_1 \\ A_2 \\ S_1 \\ S_2 \\ D \end{pmatrix}, \quad \mathbf{f}(\mathbf{u}) = \begin{pmatrix} -\tau_a^{-1} A_1 + \Omega_{NP-NP}^{eff} \tau_\sigma \tau_a^{-1} A_2 \cos kb + \Omega_R S_2 \\ -\Omega_{NP-NP}^{eff} \tau_\sigma \tau_a^{-1} A_1 \cos kb - \tau_a^{-1} A_2 - \Omega_R S_1 - \Omega_1 \\ -\tau_\sigma^{-1} S_1 - \delta \cdot S_2 - \Omega_R A_2 D \\ \delta \cdot S_1 - \tau_\sigma^{-1} S_2 + \Omega_R A_1 D + \Omega_2 D \\ 4\Omega(A_2 S_1 - A_1 S_2) - 4\Omega_2 S_2 - \tau_D^{-1}(D - D_0) \end{pmatrix}. \tag{43}$$

and the following notations were introduced $A_1 = \operatorname{Re} a_n$, $A_2 = \operatorname{Im} a_n$, $S_1 = \operatorname{Re} \sigma_n$, $S_2 = \operatorname{Im} \sigma_n$.

The equilibrium position, $\mathbf{u}_0$, of vector $\mathbf{u}$ is defined by Eqs. (33), (34), and (36). We shift vector $\mathbf{u}_0$ by $\delta \mathbf{u} \exp(\Lambda_\chi t + i\chi x)$, then after linearization we obtain

$$\delta \dot{\mathbf{u}} = \Lambda_\chi \delta \mathbf{u} = \hat{M} \delta \mathbf{u}, \tag{44}$$

where the matrix $\hat{M}$ is

$$\hat{M} = \frac{1}{\tau_a} \begin{pmatrix} -1 & \Omega_{NP-NP}^{eff} \tau_\sigma \cos \chi b & 0 & \Omega_R \tau_a & 0 \\ -\Omega_{NP-NP}^{eff} \tau_\sigma \cos \chi b & -1 & -\Omega_R \tau_a & 0 & 0 \\ 0 & -\Omega_R \tau_a D_{st} & -\tau_a \tau_\sigma^{-1} & -\tau_a \delta & -\Omega_R \tau_a A_{2st} \\ \Omega_R \tau_a D_{st} & 0 & \delta \tau_a & -\tau_a \tau_\sigma^{-1} & \Omega_R \tau_a A_{1st} \\ -4\Omega_R \tau_a S_{2st} & 4\Omega_R \tau_a S_{1st} & 4\Omega_R \tau_a A_{2st} & -4\Omega_R \tau_a A_{1st} & -\tau_a \tau_D^{-1} \end{pmatrix}. \tag{45}$$

The solution (the autowave) with a given $k$ is stable, if real parts of all eigenvalues of $\hat{M}$ are negative for any $\chi$. In Fig. 2, the dependencies of maximum values of $\operatorname{Re} \Lambda_\chi$ on $\chi$ for different values of $k$ are shown. One can see that only for $kb = \pi/2$ all values of $\operatorname{Re} \Lambda_\chi$ are negative. This confirms our assumption that only the wave arising for the minimum value of $D_{th}$ is stable.

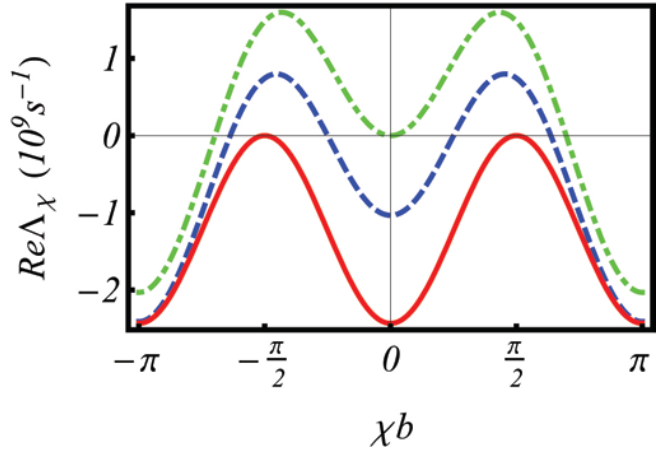

Fig. 2. (Color online) Dependencies of maximum values of $\operatorname{Re}\Lambda_\chi$ on $\chi$ for different values of $k$: dashed, dot-dashed and solid lines correspond to $kb = 0$, $\pi/4$, and $\pi/2$, respectively. $\Omega_{NP-NP} = 10^{13}\,s^{-1}$.

To make sure that for any initial condition the autowave with $k = \pi/2b$ survives, we solve nonlinear system of equations (25)-(27) numerically. We consider a chain of $N = 100$ spasers with periodic boundary conditions. We take into account the interaction with nearest neighbors only, so the first spaser interacts with the second and the hundredth ones. With the accuracy of computer simulation, we obtain that for any initial conditions, the stationary solution corresponds to one of two waves with $k = \pm\pi/2b$ and $\omega = \omega_{SP}$. The existence of two solutions is due to the inversion invariance ($x \to -x$) of the system (see Refs. 38 and references therein).

In the linear case of a NP chain, if we express the initial condition as a sum of harmonic waves, each harmonics has a fixed value of the Poynting vector. Then, each harmonics generates two waves with the same $\omega(k)$ travelling in opposite directions and having the sum of Poynting vectors equal to the initial one. Ultimately, a number of harmonic waves can travel simultaneously. The total number is determined by the initial conditions.

In our nonlinear case of a spaser chain, any initial condition evolves into a single wave which direction of propagation depends on the initial conditions. It is the total Poynting vector of the initial state that determines the direction of propagation of a single surviving wave.

Note that in the case considered in this section, there no stable solutions with $k = 0$ and $k = \pm\pi$. Thus, in a chain in which spasers are coupled via interaction of NPs only, the synchronous oscillations of spasers do not occur.

## VI. ENGAGING OF COUPLING BETWEEN NANOPARTICLES AND QUANTUM DOTS OF NEIGHBORING SPASERS

Up to this point we have assumed that there is a coupling between nearest NPs only. The QDs have been affected by the field of its own NP. Now we take into account the influence of the nearest NPs on QDs. For the sake of simplicity, the symmetric case, in which the coupling constants of interaction of a QD with neighboring NPs, $\Omega_{NP-TLS}$, are the same but they may be different from the coupling constant between the QD and the NP forming a single spaser, $\Omega_R$ (see Fig. 3).

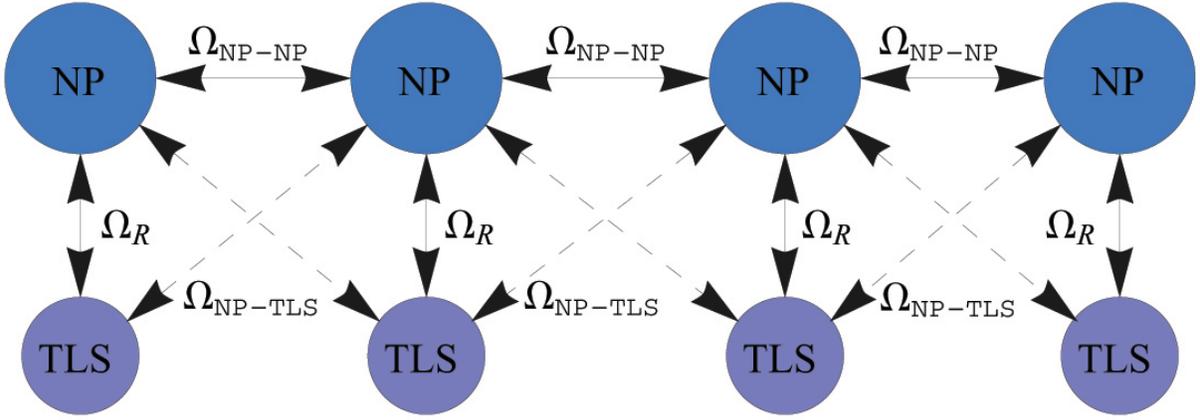

Fig. 3. (Color online) Schematic of the chain of spasers, in which each NP interacts with NPs and QDs of neighboring spasers.

The model Hamiltonian of such a system can be written as follows:

$$\hat{H} = \sum_n \left( \hbar \omega_{SP} \hat{a}_n^\dagger \hat{a}_n + \hbar \omega_{TLS} \hat{\sigma}_n^\dagger \hat{\sigma}_n + \hbar \Omega_R (\hat{a}_n^\dagger \hat{\sigma}_n + \hat{a}_n \hat{\sigma}_n^+) + 1/2 \cdot \hbar \Omega_{NP-NP} (\hat{a}_n^\dagger \hat{a}_{n-1} \right.$$
$$\left. + \hat{a}_n \hat{a}_{n-1}^\dagger + \hat{a}_n^\dagger \hat{a}_{n+1} + \hat{a}_n \hat{a}_{n+1}^\dagger) + 1/2 \cdot \hbar \Omega_{NP-TLS} (\hat{\sigma}_n^\dagger \hat{a}_{n-1} + \hat{\sigma}_n \hat{a}_{n-1}^\dagger + \hat{\sigma}_n^\dagger \hat{a}_{n+1} + \hat{\sigma}_n \hat{a}_{n+1}^\dagger) \right), \tag{46}$$

and the equation for slow-varying in time amplitudes for the $n$-th spaser are:

$$\dot{\hat{D}}_n = 2i\Omega_R (\hat{a}_n^\dagger \hat{\sigma}_n - \hat{\sigma}_n^\dagger \hat{a}_n) + 2i\Omega_{NP-TLS} (\hat{a}_{n-1}^\dagger \hat{\sigma}_n - \hat{\sigma}_n^\dagger \hat{a}_{n-1}) \tag{47}$$
$$+ 2i\Omega_{NP-TLS} (\hat{a}_{n+1}^\dagger \hat{\sigma}_n - \hat{\sigma}_n^\dagger \hat{a}_{n+1}) - \tau_D^{-1} \left( \hat{D}_n - \hat{D}_{0n} \right), \tag{48}$$

$$\dot{\hat{\sigma}}_n = \left( i\delta - \tau_\sigma^{-1} \right) \hat{\sigma}_n + i\Omega_R \hat{a}_n \hat{D}_n + i\Omega_{NP-TLS} \hat{a}_{n-1} \hat{D}_n + i\Omega_{NP-TLS} \hat{a}_{n+1} \hat{D}_n, \tag{49}$$

$$\dot{\hat{a}}_n = \left(i\Delta - \tau_a^{-1}\right)\hat{a}_n - i\Omega_R\hat{\sigma}_n - i\Omega_{NP-TLS}\hat{\sigma}_{n-1} - i\Omega_{NP-TLS}\hat{\sigma}_{n-1} - i\Omega_{NP-NP}(\hat{a}_{n-1} + \hat{a}_{n+1}).$$

If we look for solutions of Eqs. (47)-(49) in the form of the travelling wave, $a_n(t) = a_{n,k}\exp(ikx)$, $\sigma_n(t) = \sigma_{n,k}\exp(ikx)$ and $D_n(t) = D_{n,k}\exp(ikx)$, with $x = nb$, then these equations can be recast as:

$$\dot{\hat{D}}_n = 2i(\Omega_R + 2\Omega_{NP-TLS}\cos kb)(\hat{a}_n^\dagger\hat{\sigma}_n - \hat{\sigma}_n^\dagger\hat{a}_n) - \tau_D^{-1}\left(\hat{D}_n - \hat{D}_{0n}\right), \tag{50}$$

$$\dot{\hat{\sigma}}_n = \left(i\delta - \tau_\sigma^{-1}\right)\hat{\sigma}_n + i(\Omega_R + 2\Omega_{NP-TLS}\cos kb)\hat{a}_n\hat{D}_n, \tag{51}$$

$$\dot{\hat{a}}_n = \left(-i\Omega_{NP-NP}^{eff}\tau_\sigma\tau_a^{-1}\cos kb - \tau_a^{-1}\right)\hat{a}_n - i(\Omega_R + 2\Omega_{NP-TLS}\cos kb)\hat{\sigma}_n. \tag{52}$$

Comparing the system of equations (50)-(52) with Eqs. (25)-(27) one can see that accounting for the QD interaction with neighboring NPs results in the renormalization of the coupling constant, $\Omega_R$, and the frequency detuning, $\Delta$. Now, instead of $\Omega_R$ and $\Delta$, we have $\Omega_R^{nn}(k) = \Omega_R + 2\Omega_{NP-TLS}\cos kb$ and $\Delta(k) - 2\Omega_{NP}\cos kb = -\Omega_{NP-NP}^{eff}\tau_\sigma\tau_a^{-1}\cos kb$, respectively. The stationary solution of Eqs. (50)-(52) is

$$D_{st} = D_{th} = \frac{1 + \left(\Omega_{NP-NP}^{eff}\tau_\sigma\right)^2\cos^2 kb}{\left[\Omega_R^{nn}(k)\right]^2\tau_a\tau_\sigma}, \tag{53}$$

$$a_{st} = \frac{e^{i\psi}}{2}\sqrt{(D_0 - D_{th})\tau_a\tau_D^{-1}}, \tag{54}$$

$$\sigma_{st} = a_{st}\left(i\tau_a^{-1} - \left(\Omega_{NP-NP}^{eff}\tau_\sigma\right)\cos kb/\tau_a\right)/\Omega_R^{nn}(k), \tag{55}$$

Let us check the stability of this solution. The matrix $\hat{M}$, which appeared above in the linear stability analysis, modifies as:

$$\begin{pmatrix} -\tau_a^{-1} & \Omega_{NP-NP}^{eff}\tau_\sigma\cos\chi b/\tau_a & 0 & \Omega_R^{nn}(\chi) & 0 \\ -\Omega_{NP-NP}^{eff}\tau_\sigma\cos\chi b/\tau_a & -\tau_a^{-1} & -\Omega_R^{nn}(\chi) & 0 & 0 \\ 0 & -\Omega_R^{nn}(\chi)D_{st} & -\tau_\sigma^{-1} & -\delta & -\Omega_R^{nn}(\chi)A_{2st} \\ \Omega_R^{nn}(\chi)D_{st} & 0 & \delta & -\tau_\sigma^{-1} & \Omega_R^{nn}(\chi)A_{1st} \\ -4\Omega_R^{nn}(\chi)S_{2st} & 4\Omega_R^{nn}(\chi)S_{1st} & 4\Omega_R^{nn}(\chi)A_{2st} & -4\Omega_R^{nn}(\chi)A_{1st} & -\tau_D^{-1} \end{pmatrix}. \tag{56}$$

Again, the solution is stable if the real parts of all eigenvalues of the matrix $\hat{M}$ are negative. The form of the stable solution depends on the sign of the coupling constant $\Omega_R$. Let us first consider $\Omega_R > 0$. For $\Omega_{NP-TLS} = 0$, the situation is identical to the one discussed above: the stable wave has the wavenumber $kb = \pi/2$. Non-zero $\Omega_{NP-TLS} > 0$ results in a decrease of the wavenumber of the stable solution. In particular, for $\Omega_{NP-TLS} = 4 \cdot 10^{10} s^{-1}$ the stable wave has $kb = 1.2$. A further increase of the coupling constant results in a decrease of the wavenumber until it reaches zero at some value of $\Omega^*_{NP-TLS}$ (see Fig. 4). Beyond this point, the wave number remains zero when $\Omega_{NP-TLS}$ increases.

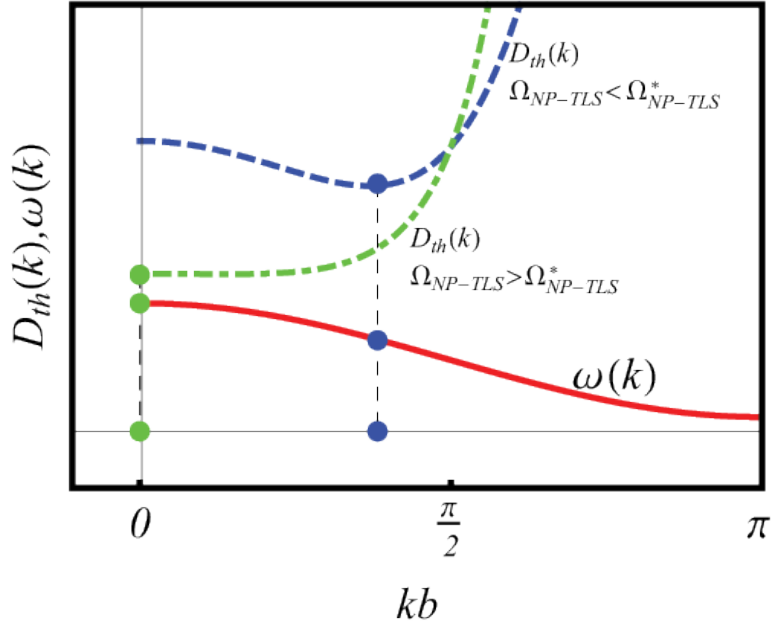

Fig. 4. (Color online) The dependence of the threshold pumping on the wavenumber $k$ for $\Omega_{NP-TLS} < \Omega^*_{NP-TLS}$ (dashed curve) and $\Omega_{NP-TLS} > \Omega^*_{NP-TLS}$ (dot-dashed curve). The solid curve shows the dependence $\omega(k)$.

The threshold value $\Omega^*_{NP-TLS}$ can be evaluated by taking into account that the pumping threshold for the stable wave has the minimum value. Eq. (53) shows that if $\omega_{SP} = \omega_{TLS}$, then

$$D_{th}(k) = \frac{1 + \left(\tau_\sigma \Omega^{eff}_{NP-NP}\right)^2 \cos^2 kb}{\left(\Omega_R + 2\Omega_{NP-TLS} \cos kb\right)^2 \tau_a \tau_\sigma} \tag{57}$$

The values of $k$ corresponding to min/max values of this expression are determined by equations:

$$\sin kb = 0, \tag{58}$$

$$\cos kb = \frac{2\Omega_{NP-TLS}}{\left(\tau_\sigma \Omega_{NP-NP}^{eff}\right)^2 \Omega_R}. \tag{59}$$

A real solution of Eq. (59) exists only if its right hand part is smaller than unity. In this case, one can show that when the inequality $2\Omega_{NP-TLS}\left(\tau_\sigma \Omega_{NP-NP}^{eff}\right)^{-2} \leq \Omega_R$ is satisfied, solutions of Eqs. (58) and (59) correspond to maximum and minimum values of $D_{th}(k)$, respectively. If $2\Omega_{NP-TLS}\left(\tau_\sigma \Omega_{NP-NP}^{eff}\right)^{-2} > \Omega_R$, the minimum of $D_{th}(k)$ is achieved for wavenumbers given by the solutions of Eq. (58). This means that

$$\Omega_{NP-TLS}^* = \frac{1}{2}\left(\tau_\sigma \Omega_{NP-NP}^{eff}\right)^2 \Omega_R. \tag{60}$$

For negative values of $\Omega_{NP-TLS}$, a similar analysis shows that for $-\Omega_{NP-TLS}^* \leq \Omega_{NP-TLS} \leq 0$ the minimum of $D_{th}(k)$ is given by the solutions of Eq. (59) and for $\Omega_{NP-TLS} \leq -\Omega_{NP-TLS}^*$ it is achieved for $k = \pi$. To summarize, in the case of $\Omega_R > 0$, the dependence of the wavenumber of the stable solution on the coupling constant $\Omega_{NP-TLS}$ is

$$kb = \begin{cases} 0, & \Omega_{NP-TLS} < -\Omega_{NP-TLS}^* \\ \pi - \cos^{-1}\left(\dfrac{2\Omega_{NP-TLS}}{\left(\tau_\sigma \Omega_{NP-NP}^{eff}\right)^2 \Omega_R}\right), & -\Omega_{NP-TLS}^* \leq \Omega_{NP-TLS} \leq \Omega_{NP-TLS}^* \\ \pi, & \Omega_{NP-TLS} > \Omega_{NP-TLS}^*. \end{cases} \tag{61}$$

Similarly, for $\Omega_R < 0$

$$kb = \begin{cases} \pi, & \Omega_{NP-TLS} < -\Omega_{NP-TLS}^* \\ \cos^{-1}\left(\dfrac{2\Omega_{NP-TLS}}{\left(\tau_\sigma \Omega_{NP-NP}^{eff}\right)^2 \Omega_R}\right), & -\Omega_{NP-TLS}^* \leq \Omega_{NP-TLS} \leq \Omega_{NP-TLS}^* \\ 0, & \Omega_{NP-TLS} > \Omega_{NP-TLS}^*. \end{cases} \tag{62}$$

The dependencies given by Eqs. (61) and (62) are shown in Figs. 4a and 4b, respectively.

Thus, depending on the value of the coupling constant between a NP and the neighboring QD, $\Omega_{NP-TLS}$, two different types of excitations in a chain of spasers arise: for $\Omega_{NP-TLS} < \Omega^*_{NP-TLS}$, a harmonic autowaves travels in the system, for $\Omega_{NP-TLS} > \Omega^*_{NP-TLS}$, there is no propagating solutions and all spasers oscillate in phase for positive spaser coupling, $\Omega_R > 0$ and if $\Omega_R < 0$, the spasers oscillate in antiphase.

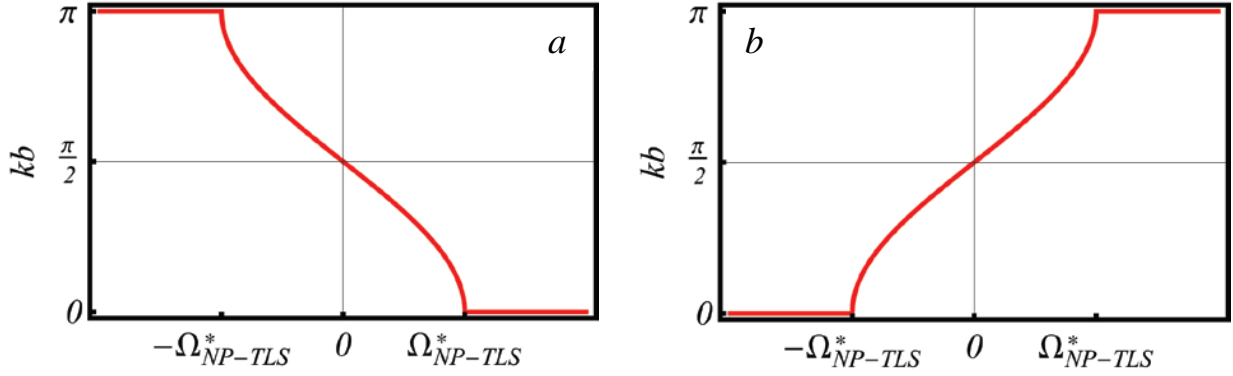

Fig. 5. The dependencies of the stable solution of Eqs. (50)-(52) for (a) $\Omega_R > 0$ and (b) $\Omega_R < 0$, respectively.

## VII. ENGAGING OF QD-QD INTERACTION

In this section, we show that taking into account the interaction of QDs belonging to neighboring spasers does not substantially change physical picture. For the sake of simplicity, we assume a symmetrical NP-QD coupling. All interactions between spasers are schematically shown in Fig. 5.

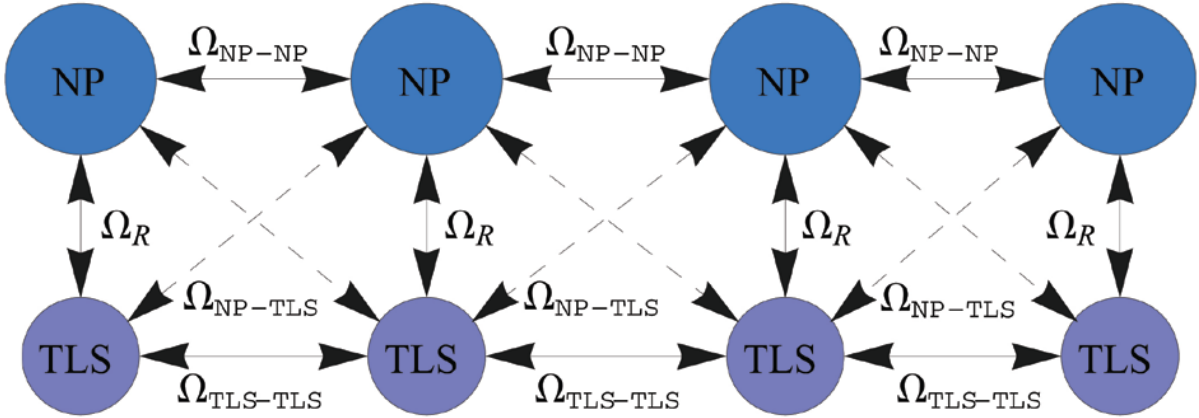

Fig. 6. (Color online) Schematic of the chain of spasers, in which each NP and QD interact with NPs and QDs of neighboring spasers.

The new Hamiltonian can be written as

$$\hat{H} = \sum_n \left[ \hbar\omega_{SP}\hat{a}_n^+\hat{a}_n + \hbar\omega_{TLS}\hat{\sigma}_n^+\hat{\sigma}_n + \hbar\Omega_R(\hat{a}_n^+\hat{\sigma}_n + \hat{a}_n\hat{\sigma}_n^+) \right.$$
$$+\frac{1}{2}\hbar\Omega_{NP-NP}(\hat{a}_n^+\hat{a}_{n-1} + \hat{a}_n\hat{a}_{n-1}^+ + \hat{a}_n^+\hat{a}_{n+1} + \hat{a}_n\hat{a}_{n+1}^+)$$
$$+\frac{1}{2}\hbar\Omega_{NP-TLS}(\hat{\sigma}_n^+\hat{a}_{n-1} + \hat{\sigma}_n\hat{a}_{n-1}^+ + \hat{\sigma}_n^+\hat{a}_{n+1} + \hat{\sigma}_n\hat{a}_{n+1}^+)$$
$$\left. +\frac{1}{2}\hbar\Omega_{TLS-TLS}(\hat{\sigma}_n^+\hat{\sigma}_{n-1} + \hat{\sigma}_n\hat{\sigma}_{n-1}^+ + \hat{\sigma}_n^+\hat{\sigma}_{n+1} + \hat{\sigma}_n\hat{\sigma}_{n+1}^+) \right], \tag{63}$$

where $\Omega_{TLS-TLS}$ is the QD-QD coupling constant. Following the procedure described in the previous sections, we arrive at the equations of motion of the spasers:

$$\dot{\hat{D}}_n = 2i(\Omega_R + 2\Omega_{NP-TLS}\cos kb)(\hat{a}_n^+\hat{\sigma}_n - \hat{a}_n\hat{\sigma}_n^+) - (\hat{D}_n - \hat{D}_{0n})\tau_D^{-1}, \tag{64}$$

$$\dot{\hat{\sigma}}_n = (i\delta - \tau_\sigma^{-1})\hat{\sigma}_n + i(\Omega_R + 2\Omega_{NP-TLS}\cos kb)\hat{a}_n\hat{D}_n + 2i\Omega_{TLS-TLS}\hat{\sigma}_n\hat{D}_n\cos kb, \tag{65}$$

$$\dot{\hat{a}}_n = -\left[\left(1 + i\Omega_{NP-NP}^{eff}\tau_\sigma\cos kb\right)\tau_a^{-1}\right]\hat{a}_n - i(\Omega_R + 2\Omega_{NP-TLS}\cos kb)\hat{\sigma}_n. \tag{66}$$

This system of equation has a stationary solution and the frequency of spasing is defined by the equation:

$$\omega^2\frac{2\Omega_{TLS-TLS}\tau_a}{(\Omega_R + 2\Omega_{NP-TLS}\cos kb)^2} + \omega\left(\tau_a + \tau_\sigma - \frac{4\Omega_{TLS-TLS}\tau_a(\omega_{NP} + 2\Omega_{NP-NP}\cos kb)}{(\Omega_R + 2\Omega_{NP-TLS}\cos kb)^2}\right) +$$
$$+\frac{2\Omega_{TLS-TLS}\tau_a\left((\omega_{NP} + 2\Omega_{NP-NP}\cos kb)^2 + \tau_a^{-2}\right)}{(\Omega_R + 2\Omega_{NP-TLS}\cos kb)^2} - \omega_{TLS}\tau_\sigma - (\omega_{SP} + 2\Omega_{NP-NP}\cos kb)\tau_a = 0. \tag{67}$$

$\Omega_{TLS-TLS}$ is much smaller than all other coupling constants because the respective dipole moments are smaller and the distance is larger. Eq. (67) has two solutions. We have to choose the one that tends to the solution of Eqs. (52) and (53) as $\Omega_{TLS-TLS}$ tends zero. For this solution, neglecting higher order terms in $\Omega_{TLS-TLS}$ we obtain the dispersion equation:

$$\omega(k) = \omega_a + \Omega_{NP-NP}^{eff}\cos kb + \frac{2\Omega_{TLS-TLS}\tau_a}{(\tau_a + \tau_\sigma)(\Omega_R + 2\Omega_{NP-TLS}\cos kb)^2}$$
$$\times\left[\left(\omega_a - \omega_{SP} - \Omega_{NP-NP}^{eff}\tau_\sigma\tau_a^{-1}\cos kb\right)^2 + \tau_a^{-2}\right]. \tag{68}$$

For the threshold value of the inversion one can obtain:

$$D_{st} = D_{th} = \frac{1 + \left(\Omega_{NP-NP}^{eff}\tau_\sigma\right)^2\cos^2 kb}{(\Omega_R + 2\Omega_{NP-TLS}\cos kb)^2\tau_a\tau_\sigma}, \tag{69}$$

where $\Delta = \omega(k) - \omega_{NP}$. Note that both $\omega(k)$ and $D_{th}(k)$ depend on $\cos kb$ only. The result of the differentiation of Eq. (69) with respect to $k$ is $\sin kb$ multiplied by a function that has no zeros. Therefore, extremums of $D_{th}(k)$ are achieved for $kb = 0$ and $kb = \pi$. Whether we have maximum or minimum of $D_{th}(k)$ depends on the sign of the second derivative, which for small values of $\Omega_{TLS-TLS}$ is the same as in Sec. VI. Thus, when a QD-QD coupling is "turned on," the wavenumbers of the stable solution are still either $kb = 0$ or $kb = \pi$ depending upon the sign of $\Omega_R$. Of course, the threshold value of $\Omega_{NP-TLS}^*$ is different than the one given by Eq. (60).

## VIII. DISCUSSION

In this paper, we have studied excitations in a chain of interacting spasers. We have shown that depending on the strength of the interaction between a QD and the nearest NP, either a synchronized oscillation of all the spasers or a harmonic autowave travelling along the chain may realize. Thus, the pumped QD may either excite its own spasers so that all spasers are synchronized or cooperating with the other QDs, the pumped QD may excite a plasmonic wave traveling along the chain. This is the wave of the NP polarization which dispersion equation is similar to the one predicted in Refs. 2 and 11 for linear systems. Unlike the general case of a wave propagating in a nonlinear lattice,[38] the nonlinear character of the spasers' response to an external field results neither in soliton nor in kink solutions. Rather, this response is a perfectly harmonic wave. However, unlike harmonic waves in linear systems, in a chain of spasers, (*i*) the wave has a fixed value of the wavenumber, which is determined by the minimum value of the pumping threshold values of the coupling constants, (*ii*) its amplitude also has a fixed value, which is determined by the pumping strength, and (*iii*) its propagation direction is determined by the initial conditions.

## ACKNOWLEDGEMENTS

The authors are indebted to J. B. Pendry, who drew our attention to the possibility of the stationary excitation at the chain of spasers in the form of the travelling wave. This work was supported by RFBR Grants Nos. 10-02-91750, 10-02-92115, 11-02-92475 and 12-02-01093 and by a PSC-CUNY grant.